\documentclass[]{raa}                  
\usepackage{graphicx,times}             
\input{epsf.sty}                        
\input{psfig.sty}                       

\begin{document}

   \title{Magnetic Gradient: A Natural Driver of Solar Eruptions}

   \volnopage{Vol.00 (2019) No.0, 000--000}      
   \setcounter{page}{1}          

   \author{Baolin Tan \inst{1, 2} \and Yan Yan \inst{1} \and Ting Li \inst{1} \and Yin Zhang \inst{1} \and Xingyao Chen \inst{1,2}}
   \institute{Key Laboratory of Solar Activity, National Astronomical
Observatories of Chinese Academy of Sciences, Beijing 100012,
China, {\it bltan@nao.cas.cn}\\
   \and School of Astronomy and Space Science, University of Chinese Academy of Sciences, Beijing 100049, China\\ }

   \date{Received~~2019 month day; accepted~~2019~~month day}

\abstract{It is well-known that there is a gradient, there will drive a flow inevitably. For example, a density-gradient may drive a diffusion flow, an electrical potential-gradient may drive an electric current in plasmas, etc. Then, what will be driven when a magnetic-gradient occurs in solar atmospheric plasmas? Considering the ubiquitous distribution of magnetic-gradient in solar plasma loops, this work demonstrates that magnetic-gradient pumping (MGP) mechanism is valid even in the partial ionized solar photosphere, chromosphere as well as in the corona. It drives energetic particle flows which carry and convey kinetic energy from the underlying atmosphere to move upwards, accumulate around the looptop and increase there temperature and pressure, and finally lead to eruptions around the looptop by triggering ballooning instabilities. This mechanism may explain the formation of the observing hot cusp-structures above flaring loops in most preflare phases, therefore, the magnetic-gradient should be a natural driver of solar eruptions. Furthermore, we may also apply to understand many other astrophysical phenomena, such as the temperature distribution above sunspots, the formation of solar plasma jets, type-II spicule, and fast solar wind above coronal holes, as well as the fast plasma jets related to white dwarfs, neutron stars and black holes.
\keywords{Sun: eruptions -- Sun: corona -- Sun: flare} }

   \authorrunning{Tan, Yan, Li, Zhang, and Chen}            
   \titlerunning{Triggering Mechanism of Solar Eruptions}  

   \maketitle
\section{Introduction}
\label{sect:intro}

Solar eruptions, including solar flares, coronal mass ejections (CMEs), and various scales of plasma jets release a great amount of energy (up to $10^{25}$ J in tens of minutes in a typical X-class flare), eject fast hot plasma flows (up to 1000 km s$^{-1}$), and emit a great number of energetic particles into the interplanetary space, and produce great impacts on the terrestrial environment. Although studies of solar eruptions have been done for more than one century, there are still many big questions, including what powers the eruptions? What is the primary trigger? etc. Answers of these questions may help us to better predict when, where, and how solar eruptions occur, and avoid their damages as soon as possible.

Observations show that most solar eruptions take place in active regions which composed of many plasma loops (Somov 1989, Shibata 1999, Wang et al. 2002, Tan et al. 2006) with scales from several hundred km to beyond one million km, and stretching from the photosphere, via chromosphere to the high corona (Bray 1991, Hudson 2011). Many models proposed that magnetic reconnection could release magnetic energy, accelerate particles and heat plasmas (Lin 2003, Schrijver 2009, Chen 2011). Before the onset of magnetic reconnection, the active region has stored enough free energy by twisting or shearing motions near the photosphere (Ishii 2000, Fang 2012), or braiding of plasma loops by continuous footpoint motions (Cirtain 2013, Tiwari 2014), or magnetic flux emergence (Liu 2006) and other motions. However, what trigger these eruptions is still debated (Hu et al. 1995, Forbes 2000, Schrijver 2009, Shibata et al. 2011, Kusano et al. 2012, Aulanier 2014, Sun et al. 2015, Jiang et al. 2016, Wyper et al. 2017).

\begin{figure}[ht]
\begin{center}
   \includegraphics[width=9 cm]{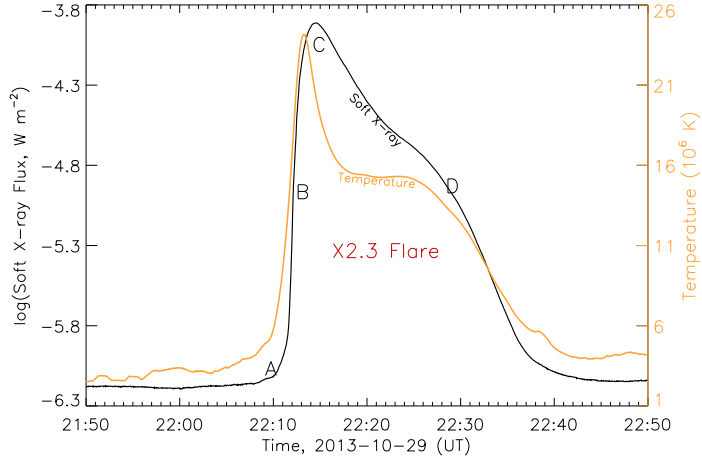}
\caption{Eruptive process of an X2.3 flare observed at SXR of 1 - 8 \AA~ by GOES satellites on 2013 October 29. The orange curve shows the temperature in the flaring source region derived from the observed SXR emission fluxes at two wavelengths: 1 - 8 \AA~ and 0.5 - 4 \AA.}
\end{center}
\end{figure}

Fig.1 presents a typical solar flare observed by GOES soft X-ray (SXR) telescope. The black curve presents the SXR emission at 1 - 8 \AA, and the orange curve shows the temperature in flaring region derived from SXR emission at two wavelength band (1 - 8 \AA~ \& 0.5 - 4 \AA) (Thomas et al. 1985, White et al. 2005). The whole process can be plotted into 4 phases: slow increase (A), fast increase (B), maximum (C), and gradual decrease (D). The slow increase lasts for more than 10 min before the flare start with slowly temperature increasing in source region. Then, the emission increases rapidly at 4 orders of magnitude in about 3 min and the temperature increases from about $4\times10^{6}$ K up to $2.4\times10^{7}$ K in about 2 min (B). After the maximum (C), both the emission flux and temperature decrease gradually in more than 20 min. Here, we noted that the temperature reaches its maximum 1 - 2 min prior to the SXR maximum. If the flare eruption is generated from magnetic reconnection, then when did the reconnection start? Does it start in phase A or phase B? If it occurred in phase A, why the temperature increases slowly? If it occurred in phase B, why the temperature increases before this phase? Obviously there must have some heating processes in source region in phase A. Then, what heat the source region during phase A? and what is the physical connection between the heating process and magnetic reconnection? Many models have been proposed to explain the triggering mechanism of solar eruptions (Forbes et al. 2006, Kusano et al. 2012, Schmieder et al. 2013, Aulanier, 2014). However, two basic questions still remain open: how does the energy accumulate in the source region before the onset of eruptions? which drive and trigger the plasma loops to erupt?

It is well-known that there is a gradient, there will drive a flow inevitably. For example, an electric potential-gradient ($\nabla U$) will drive an electric current in plasmas: $j=-\sigma\cdot\nabla U$ ($\sigma$ is the conductivity), a density-gradient ($\nabla n$) may drive a diffusion flow: $\Gamma=-D\cdot\nabla n$ ($D$ is the diffusion coefficient), and a temperature gradient ($\nabla T$) may drive a heat flow: $q=-K\cdot\nabla T$ ($K$ is the coefficient of heat conduction), etc. Then, what flow will be driven by a magnetic-gradient ($\nabla B$) in solar plasma loops?

Considering the ubiquitous magnetic gradient in solar atmosphere, this work proposed that magnetic-gradient in solar plasma loops may drive an energetic particle flow which carry and convey kinetic energy from the solar lower atmosphere to move upwards, accumulate around the looptop and increase the temperature, and finally trigger a violent eruption. Section 2 introduces the magnetic-gradients in solar atmosphere. Section 3 discuss the primary processes driven by magnetic-gradient force. Section 4 is the applications to other astrophysical processes, such as the formation of the cold of sunspot near photosphere and the hot above sunspot, plasma jets, type-II spicule, and the fast solar wind above coronal holes, etc. Finally, conclusions are summarized in Section 5.

\section{Magnetic-gradient in solar atmosphere}

The solar atmosphere is always filled with many magnetized plasma loops in lengths from several thousand km to beyond one million km. Some of them connect two opposite magnetic polarities in an active region, while some of them may even connect different active regions, and the solar global magnetic field may connect both solar poles. Because generally the plasma is frozen in magnetic field, these loops are shaped the main structure of solar atmosphere with different scales of lengths and heights.

In each solar plasma loop, the magnetic field around the footpoint near the photosphere is strongest and decreases generally with the increasing height above the solar surface. Therefore, there is magnetic-gradient ($\nabla B$) from the photosphere to the corona, and the direction of magnetic-gradient is downward. So far, we have no reliable direct measurement of the magnetic field and the gradient in the chromosphere and corona. We may indirectly estimate them above active regions from a fitted expression obtained by Dulk \& MeClean (1978):

\begin{equation}
B=0.5(\frac{r}{R_{s}}-1)^{-\frac{3}{2}}=0.5(\frac{R_{s}}{h})^{\frac{3}{2}}.
\end{equation}

The unit of magnetic field strength $B$ is Gs. $R_{s}$ is the solar photospheric radius. $h$ is the height above the photosphere. From Equation (1), the vertical magnetic-gradient can be derived,

\begin{equation}
\frac{dB}{dh}\approx-1.08\times10^{-9}(\frac{R_{s}}{h})^{\frac{5}{2}}.
\end{equation}

From Equation (1) and (2), we may estimate that the magnetic field strength and magnetic-gradient are about 102 Gs and -7.7$\times10^{-6}$ Gs m$^{-1}$ at height of 2$\times10^{4}$ km, 26 Gs and -7.8$\times10^{-7}$ Gs m$^{-1}$ at height of 5$\times10^{4}$ km, 9 Gs and - 1.4$\times10^{-7}$ Gs m$^{-1}$ at height of 1$\times10^{5}$ km, 3 Gs and - 2.4$\times10^{-8}$ Gs m$^{-1}$ at height of 2$\times10^{5}$ km, respectively. Some roughly observations show that the magnetic field is about 1000 G around the footpoint, 10-250 G at height about 5$\times10^{4}$ km, and 5-10 G at height about 2$\times10^{5}$ km, respectively. Accordingly, the magnetic-gradient is about -4$\times10^{-7}$ G m$^{-1}$ at height of 5$\times10^{4}$ km, and -2$\times10^{-9}$ G m$^{-1}$ at the height of 2$\times10^{5}$ km, respectively (Gelfreikh et al. 1997, Mathew \& Ambastha 2000, Cui et al. 2007, Joshi et al. 2017).

Many practices show that Equations (1) and (2) is valid only in the range of $h=0.02 - 10 R_{s}$ above solar surface with uncertainty $\leq 30\%$. In the lower solar atmosphere, especially in the photosphere, chromosphere and lower corona with height $h<10$ Mm, we have to obtain the magnetic field and its gradient from modeling extrapolations. All of the above approaches show that the magnetic-gradient near the footpoint gets its maximum, then decreases rapidly with height and diminishes in the high corona.

\section{Magnetic-gradient Driving Processes of Solar Eruptions}

\subsection{Principle of Magnetic-gradient Pumping Mechanism}

Tan (2014) proposed that a charged particle may have the following balance in solar atmospheric plasma with slowly-varying inhomogeneous magnetic field,

\begin{equation}
F_{t}=F_{m}+mg(h).
\end{equation}

Here, $mg(h)$ is the solar gravitational force at height $h$ above the solar surface. $m$ is the mass of the charged particle, $g(h)=\frac{GM_{s}}{(R_{s}+h)^{2}}$ is the solar gravitational acceleration at $h$. $M_{s}$ and $R_{s}$ are the mass and photospheric radius of the Sun, respectively. $F_{m}$ is the magnetic-gradient force which can be expressed as,

\begin{equation}
F_{m}=-\mu\cdot\nabla B=-G_{B}\cdot\epsilon_{t}.
\end{equation}

Here, $\mu=\frac{\frac{1}{2}mv_{t}^{2}}{B}=\frac{\epsilon_{t}}{B}$ is the magnetic-moment which is approximately an invariance in the slowly-varying inhomogeneous magnetized plasmas. $\nabla B$ is the magnetic-gradient along the magnetic field lines, $v_{t}$ is the transverse velocity, and $\epsilon_{t}$ is the transverse kinetic energy. $G_{B}=\nabla B/B$ is the relative magnetic-gradient. $L_{B}=\frac{1}{G_{B}}$ is the magnetic field scale length.

Because $F_{m}\propto-G_{B}$, the magnetic-gradient ($G_{B}$) plays an effective force on the charged particles and drive it to get away from the strong magnetic field region to the weak field region. In solar conditions, the relative magnetic-gradient $G_{B}$ is nearly constant at certain place in plasma loop, $F_{m}$ only changes with respect to the particle's transverse kinetic energy: $F_{m}\propto\epsilon_{t}$. The higher the kinetic energy of a particle, it will get a stronger $F_{m}$, and get away from strong field region faster than the low energy particles. The energetic particles are picked up by the magnetic-gradient force from the underlying thermal plasma with strong magnetic field, transported to move upwards, accumulate in the high plasma with weak magnetic field in solar plasma loops. The plasma loops act as a pumper driving energetic particles (similar as water) to move upwards and form an energetic particle flow. Therefore, this process is called magnetic-gradient-pumping (MGP) mechanism (Tan 2014).

However, Equations (3) and (4) are derived under conditions of collision-free plasmas and slow variation of magnetic field in time and space, which may ensure that a charged particle can finish at least one more cycles gyrating to the magnetic field before it collides with other particles. In this case, the magnetic-moment is approximately conserved. These conditions are satisfied in the solar corona and upper chromosphere. But the photosphere and lower chromosphere are partial ionized and frequent collisions for their low temperature and high density. It seems that Equations (3) and (4) are possibly not valid here. Just because of this doubt, there is little response since Tan (2014) proposed the MGP mechanism to explain the mystery of coronal heating.

It is possibly inspirational to compare the collision timescales ($t_{c}$) and the magnetic cyclotron period ($t_{mc}$) in the solar photosphere and chromosphere.

(1) Collision timescale ($t_{c}$). In solar photosphere and chromosphere, because of the weakly partial ionization, the dominated collision mainly occur between the charged particles and the neutral hydrogen atoms. the collision timescale can be estimated by $t_{c}(ia)\approx\frac{1}{\pi r^{2}vn_{n}}$. Here, $r\approx5.3\times10^{-11}$ m is the radius of hydrogen atom, $n_{n}$ is the density of hydrogen atoms, $v\approx(\frac{k_{B}T}{m})^{\frac{1}{2}}$ is the averaged speed of the charged particles (mainly protons). Approximately, $t_{c}(ia)\approx1.25\times10^{18}\frac{1}{n_{n}T^{1/2}}$. The other collision timescale is occurred among ions, $t_{c}(ii)\approx4.64\times10^{5}\frac{T^{3/2}}{n_{i}}$. $n_{i}$ is the density of ions.

(2) Magnetic cyclotron period ($t_{mc}$). The proton's magnetic cyclotron period can be simply expressed as $t_{mc}\approx6.7\times10^{-8}B^{-1}$. Here, the unit of magnetic field B is Gauss.

\begin{table}
\begin{center}
 \caption[]{The comparison between the collision timescales and the magnetic cyclotron periods in solar atmosphere.}
 \begin{tabular}{ccccccccccc}
  \hline\noalign{\smallskip}
 Parameter           & Photosphere           & Chromosphere         &  Corona             \\\hline\noalign{\smallskip}
   T (K)             &    5450               &   10800              & 447000               \\
 $n_{n}$ (m$^{-3}$)  & 6.880$\times10^{22}$  & 9.136$\times10^{16}$ & 2.137$\times10^{15}$ \\
 $n_{i}$ (m$^{-3}$)  & 1.065$\times10^{19}$  & 7.259$\times10^{16}$ & 2.567$\times10^{15}$ \\
   B (Gs)            &      500              &    100               &    20            \\
 $t_{c}(ia)$ (s)     & 2.46$\times10^{-7}$   &    0.13              &    0.87         \\
 $t_{c}(ii)$ (s)     & 1.75$\times10^{-8}$   & 7.2$\times10^{-6}$   & 5.4$\times10^{-2}$  \\
 $t_{mc}$ (s)        & 1.34$\times10^{-10}$  & 6.7$\times10^{-10}$  & 3.4$\times10^{-9}$  \\
\noalign{\smallskip}\hline
\end{tabular}
\end{center}
\tablecomments{0.96\textwidth}{The data of temperature, hydrogen density and ion density are cited from Vernazza et al. (1981).}
\end{table}

Table 1 lists the comparison between the typical collision timescale and the magnetic cyclotron periods in solar atmosphere. Here the data of temperature, magnetic field strength, hydrogen density and ion density in the photosphere, chromosphere, and corona are cited from Vernazza et al. (1981). The comparison shows that the magnetic cyclotron periods are much shorter than the collision timescale: $t_{mc}\ll t_{c}$. Even in the photosphere, the proton's magnetic cyclotron period is shorter at least 100 times than the collision timescale. This means that a proton can gyrate at least more than 100 cycles before it collides with an atom or more than 1000 cycles before it collides with a proton. This fact indicates that Equation (3) and (4) are valid even in the photosphere. There is enough time for MGP mechanism works in solar plasma loops.

Although the collision timescale is much longer than the proton's magnetic cyclotron period, it is still much shorter than the lifetime of solar plasma loops ($t_{d}$) which is generally from many hours to several days ($t_{d}>10^{5}$ s). $t_{d}\gg t_{c}$ means that there are enough time for pumping particles to transform their kinetic energy into the thermal energy by collisions in the solar plasma loops. Therefore, the MGP process might play a significant role for heating the upper atmosphere even in the solar photosphere, chromosphere as well as in the corona.

Actually, because of the curvature of magnetic field lines in a closed plasma loop (Fig. 2), Equation (3) should be modified into,

\begin{equation}
F_{t}=-G_{B}\cdot\epsilon_{t}\cos\theta+mg(h).
\end{equation}

When $F_{t}>0$, then $\epsilon_{t}>\frac{mg(h)}{G_{B}\cdot\cos\theta}$, the particle will get rid of the confinement of the solar gravitation force and fly upward along the plasma loop, called escaping particle, or pumping particles. When $F_{t}<0$, then $\epsilon_{t}<\frac{mg(h)}{G_{B}\cdot\cos\theta}$, the particle will be confined in the lower region and not move upward, called confined particle. The threshold of the particle's transverse kinetic energy is called starting energy, expressed as,

\begin{equation}
\epsilon_{t0}=\frac{mg(h)}{G_{B}\cdot\cos\theta}=\frac{mg(h)}{\cos\theta}L_{B}
\end{equation}

Obviously, the solar gravitational force plays a key role in the MGP model.

\begin{figure}[ht]   
\begin{center}
   \includegraphics[width=7.5 cm, height=5.5 cm]{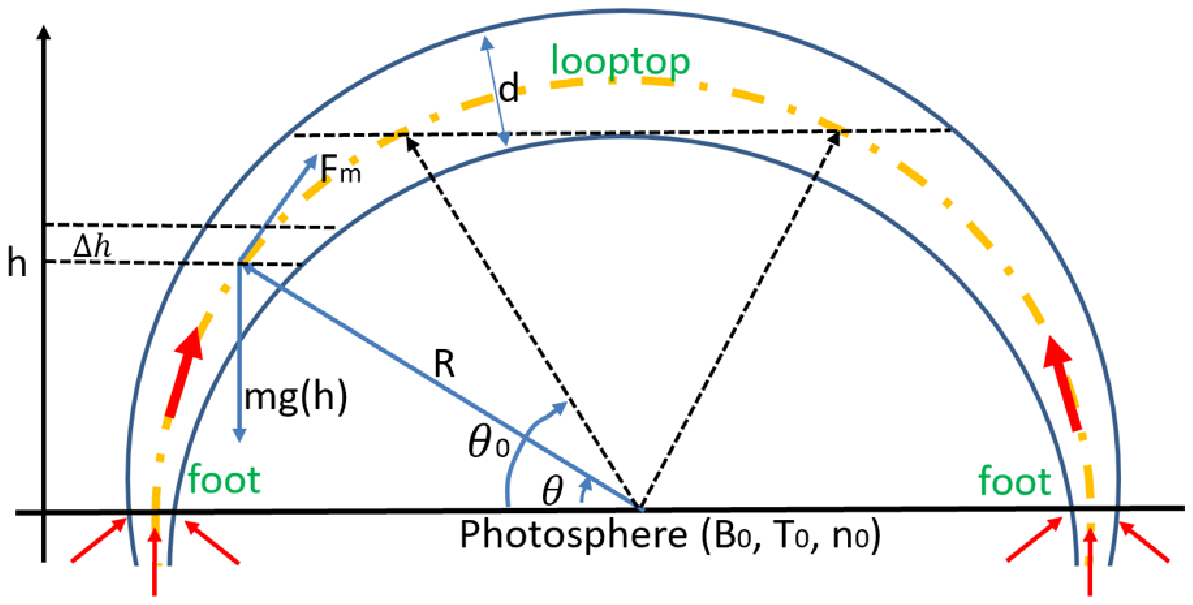}
   \includegraphics[width=7.2 cm, height=5.5 cm]{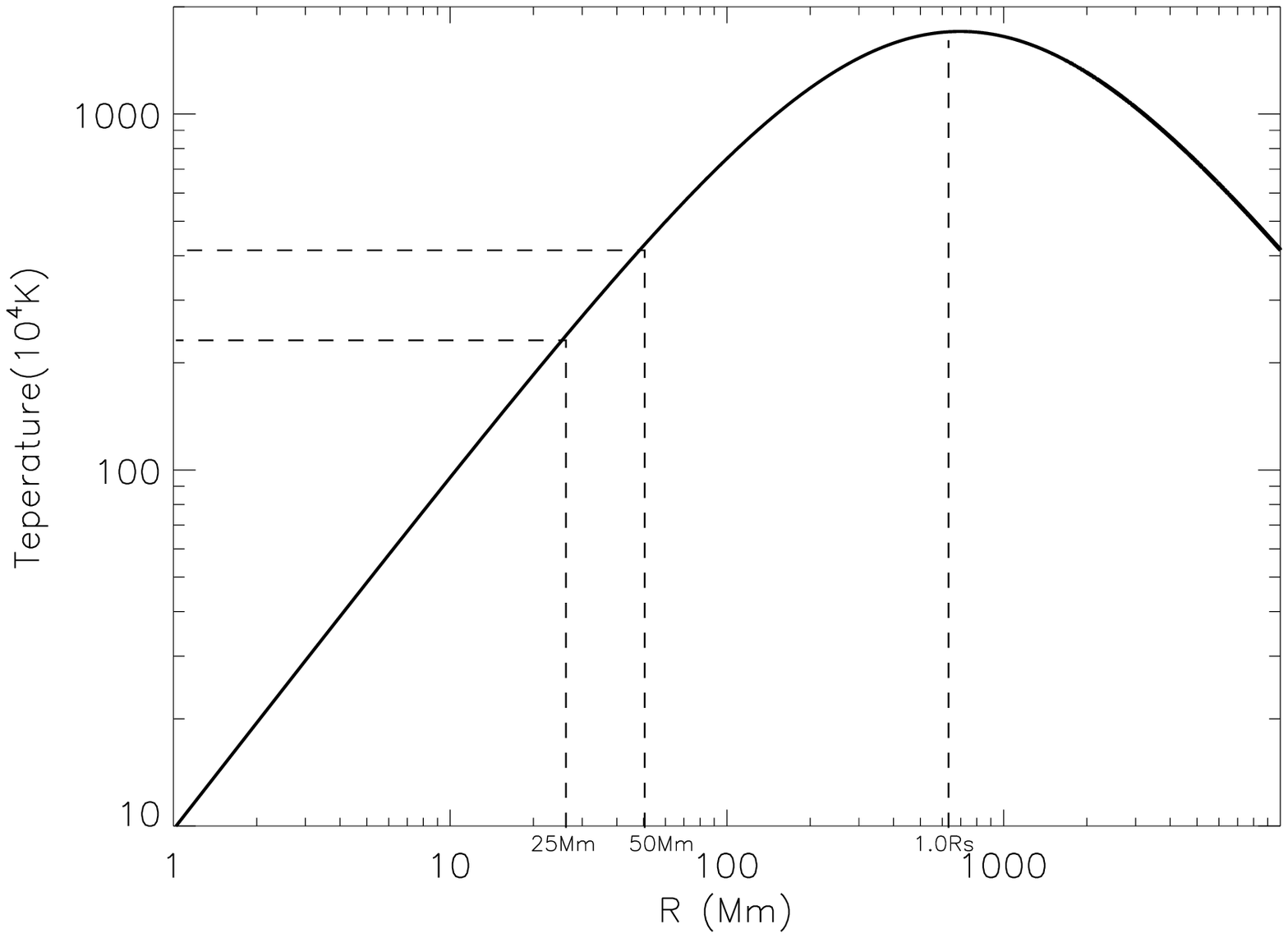}
\caption{The schematic diagram of the MGP mechanism and temperature distribution in solar magnetic plasma loop. The big red arrows show the motion of escaping energetic particles. The small red arrows show the motion of energetic particles compensating from the inner part of the Sun driven by the convection motion. The right panel shows the temperature around the looptop with radius of R.}
\end{center}
\end{figure}

The starting energy is a function of the height ($h$) above the solar surface. For a symmetric semicircle plasma loop (Fig. 2), using Equation (1) and (2) approximately, the starting energy can be expressed as,

\begin{equation}
\epsilon_{t0}(h)\approx1.9\times10^{-6}\frac{h}{(1+\frac{h}{R_{s}})^{2}\cos\theta} (eV).
\end{equation}

At each height $h$ below looptop of the plasma loop, the kinetic energy of proton should be $\epsilon_{t0}(h)$ (as discussed in Tan 2014, protons play dominated role in the solar process). Any protons with kinetic energy of $\epsilon_{t}>\epsilon_{t0}(h)$ should be triggered to fly away to higher place, while the protons with kinetic energy of $\epsilon_{t}<\epsilon_{t0}(h)$ will be confined in the lower place. Therefore, the starting energy $\epsilon_{t0}(h)$ is a monotonic function of height $h$, and the corresponding temperature at height $h$ can be expressed as,

\begin{equation}
T(h)\approx2.2\times10^{-2}\frac{h}{(1+\frac{h}{R_{s}})^{2}\cos\theta}~(K).
\end{equation}

Solar plasma loops always have specific widths $d$ and the looptop occupies a large area (Fig.2). In a symmetric loop, the looptop region can be defined as $\theta\geq\theta_{0}$. Here, $\cos\theta_{0}=\sqrt{1-(1-\frac{d}{2R})^{2}}\approx(\frac{d}{R})^{1/2}$. The looptop traps the particles with kinetic energy $\epsilon_{t0}>1.9\times10^{-6}\frac{h}{(1+\frac{h}{R_{s}})^{2}}(\frac{R}{d})^{1/2}$ (eV), and the temperature should be $T_{top}>2.2\times10^{-2}\frac{R}{(1+\frac{R}{R_{s}})^{2}}(\frac{R}{d})^{1/2}$~(K). Usually, the ratio of $\frac{R}{d}$ is about 20 (Bray et al. 1991), then, the temperature would exceed 2.2$\times10^{6}$ K around the looptop with radius of 25 Mm, and exceed 4.0$\times10^{6}$ K with loop radius of 50 Mm. It may reach to a maximum (exceeds $10^{7}$ K) when the loop radius is at 1.0 $R_{s}$ (right panel of Fig.2).

The pumping particles are picked up by the magnetic-gradient force from the underlying thermal plasma, transported to move upwards, accumulate in the high corona, and finally increase the averaged particle kinetic energy of the coronal plasma. The plasma loops act as a pumper driving energetic escaping particles (similar as water) to move upwards and form an energetic particle flow. Because temperature is a measurement of particles' averaged kinetic energy in a plasma volume with thermal equilibrium, the above process consequently increases the temperature of the corona, equivalently heat the corona. Therefore, this process is called magnetic-gradient-pumping (MGP) mechanism (Tan 2014). Fig. 3 shows the MGP process and the formation of energetic particle upflow. The red circles represent the pumping particles while the black circles indicate the confined particles.

\begin{figure}[ht]   
\begin{center}
   \includegraphics[width=10 cm]{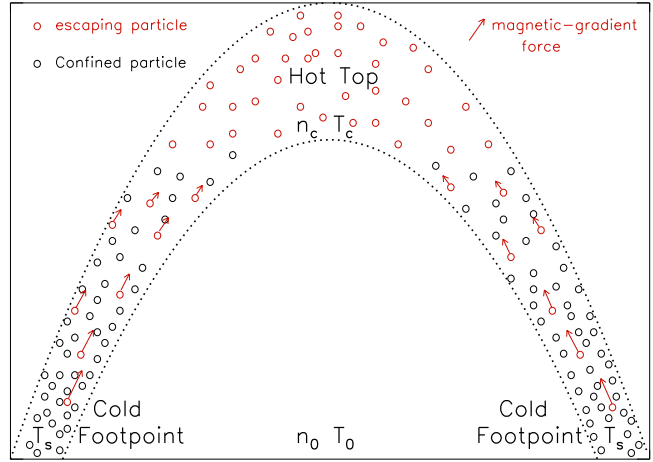}
\caption{The schematic diagram of the magnetic-gradient-pumping process and the formation of energetic particle flow in a plasma loop. Red circles represent the escaping particles driven by magnetic-gradient force, and the black circles represent the confined particles, red arrows indicate the energetic particle flow.}
\end{center}
\end{figure}

The above estimations are roughly approximation based on the simple assumptions of magnetic field in Equation (1) and (2) and the symmetric semicircle plasma loop (Fig. 2). The real solar conditions will be much more complex than the above regime. However, the underlying physical process and the results should be very similar.

Additionally, we need to answer another question: are there enough charged particles for magnetic-gradient pumping to move upward in the weakly ionized photosphere?

At first, the degree of ionization ($a_{i}$) may indicate that how much charged particles in the weakly ionized plasma, it can be calculated from the Saha's Equation:

\begin{equation}
a_{i}\approx4.9\times10^{10}\frac{T^{3/4}}{\sqrt{n_{n}}}exp(-\frac{U_{i}}{2k_{B}T}).
\end{equation}

$U_{i}$ is the ionized potential energy of hydrogen atom, $U_{i}=13.6$ eV.

Then, we may calculate the fraction of the magnetic-gradient pumping particles (pumping rate, $a_{pu}$) by the following integral,

\begin{equation}
a_{pu}=\int_{\epsilon_{t0}}^{\infty}
f(\epsilon_{k})d\epsilon_{k}.
\end{equation}

$f(\epsilon_{k})$ is the distribution function of particles in the photosphere. Generally we may suppose that it is a Maxwell distributing function: $f(\epsilon_{k})=\frac{\epsilon_{k}}{(k_{B}T)^{2}}exp(-\frac{\epsilon_{k}}{k_{B}T})$. Then, the pumping rate can be expressed as,

\begin{equation}
a_{pu}=\int_{\epsilon_{t0}}^{\infty}
\frac{\epsilon_{k}}{(k_{B}T)^{2}}exp(-\frac{\epsilon_{k}}{k_{B}T})d\epsilon_{k}.
\end{equation}

The starting energy $\epsilon_{t0}$ is very crucial in above calculation. Here, we can not obtain $\epsilon_{t0}$ directly from Equation (7) in the solar photosphere. Because Equations (1) and (2) and (7) are valid only when the height is at the range of 0.02 - 10 $R_{s}$, they are not valid near the photosphere for their magnetic field is mainly vertical to the solar surface. The magnetic modeling and extrapolations show that the magnetic field scale length near solar surface is about $L_{B}\sim2\times10^{6}$m, and then $\epsilon_{t0}\approx5.7$ eV. The calculating results of $a_{i}$ and $a_{up}$ are listed in Table 2. It is showed that $a_{i}>a_{up}$ even at the minimum temperature region of the solar photosphere. This fact implies that there are enough charged particles for the MGP process even in the weakly ionized photosphere. Actually, we may think about this process from another way: the density of the photosphere is generally at about $10^{22}-10^{23}$ m$^{-3}$, the temperature is about 5780-6400 K, and the degree of ionization is at the order of $10^{-4}$. That means the density of the charged gas is still at least at the order of $10^{18}-10^{19}$ m$^{-3}$ in the photosphere, and this is much higher than the density of the hot corona ($10^{14}-10^{16}$ m$^{-3}$). This means that the photosphere can provide enough charged gas for the upper hot atmosphere, needn't extra heating processes.

\begin{table}
\begin{center}
 \caption[]{The comparison between the degree of ionization and the fraction of the magnetic-gradient pumping particles in the photosphere.}
 \begin{tabular}{ccccccccccc}
  \hline\noalign{\smallskip}
   h (km) &  T (k) & $n_{n}$ (m$^{-3}$)    &  $a_{i}$           &   $a_{up}$         \\\hline\noalign{\smallskip}
   0      &  6420  & 1.168$\times10^{23}$  & 4.8$\times10^{-4}$ & 3.6$\times10^{-4}$ \\
   500    &  4440  & 2.483$\times10^{21}$  & 1.1$\times10^{-5}$ & 5.0$\times10^{-6}$ \\
   855    &  5890  & 9.996$\times10^{19}$  & 5.1$\times10^{-3}$ & 1.5$\times10^{-4}$ \\
   1280   &  6510  & 5.723$\times10^{18}$  & 8.2$\times10^{-2}$ & 4.1$\times10^{-4}$ \\
   1515   &  6740  & 1.494$\times10^{18}$  &      0.25          & 5.6$\times10^{-4}$ \\
\noalign{\smallskip}\hline
\end{tabular}
\end{center}
\tablecomments{0.96\textwidth}{The data of temperature and hydrogen density at different height $h$ above the solar surface are cited from Vernazza et al. (1981).}
\end{table}

The motion of the escaping particles forms a natural upflow of energetic particles in open magnetic field configurations. This energetic upflow may explain the formation of some solar ejection phenomena, which will be presented in Section 4.

\subsection{MGP Triggering Mechanism in Solar Eruptions}

In solar plasma loops, the pumping particles move upward driven by the magnetic-gradient force from both footpoints. The flying timescale ($t_{f}$) of pumping particles to fly upward from the footpoint via the plasma loop and reach to the looptop can be estimated by,

\begin{equation}
t_{f}\approx\frac{R}{v_{\|}}\approx\frac{3m_{i}}{2\epsilon_{t0}}\approx1.93\times10^{-7}R(1+\frac{h}{R_{s}})\sqrt{\frac{R_{s}}{h}}.
\end{equation}

For a loop with radius of R=25 Mm, $t_{f}\sim 26$ s, and $t_{f}\sim 38$ s in a loop with radius of R=50 Mm. Typically, $t_{f}\sim10-100$ s in most cases, it is much shorter than the lifetime of solar plasma loop: $t_{f}\ll t_{d}$. This fact implies that the plasma loop has enough time to be heated by the pumping energetic particles from the MGP process.

The pumped energetic particles pile up and accumulate around the looptop, and this will result in increasing of the particle density ($n_{e}$) and temperature ($T_{e}$). It is equivalent to a heating process. Consequently, the plasma thermal pressure ($p_{t}=k_{B}n_{e}T_{e}$) increases, and the plasma parameter $\beta=\frac{p_{t}}{p_{m}}$ will also increase. Here, $p_{m}=\frac{B^{2}}{2\mu_{0}}$ is the magnetic pressure. Finally, when $\beta$ exceeds a critical value $\beta_{c}$ the magnetic pressure cannot balance the expansion of the plasma thermal pressure, and the plasma loop will loss its equilibrium, produce a ballooning instability around the looptop, break away from the confinement of magnetic field, let out energetic particles and kinetic energy, and finally lead to violent magnetic eruptions.

In practice, the critical plasma beta $\beta_{c}$ is very small ($\beta_{c}\ll1$, means $p_{t}\ll p_{m}$). The $\beta_{c}$ value depends on the boundary conditions (Haas \& Thomas 1973, Greenwald et al 1988, Greenwald, 2002), including the radii of the magnetic loop and its cross-section, the distributions of plasma density, magnetic field, and current density, etc. The Tokmak experimental results show $\beta_{c}<0.1$ (Inverarity \& Priest 1996, Tsap et al. 2008, Katsuro-Hopkins et al. 2010). From the critical plasma beta, the threshold parameter becomes:

\begin{equation}
M=n_{m}T_{e}=\frac{B^{2}}{2\mu_{0}k_{B}}\beta_{c}.
\end{equation}

$M$ describes a critical status of magnetized plasma loops. $n_{m}$ is the density limit of the ballooning instability. The increases of either plasma density or temperature will make the plasma approaching the critical status, and excite ballooning instability in the plasma loop. Tsap et al. (2008) investigated the excitation of the ballooning instability in a coronal flaring loop under the framework of ideal MHD and found that ballooning instability would be excited when $\beta_{c}\approx\frac{d}{R}$. $R$ is the radius of loop's curvature. Generally, $\frac{d}{R}\approx0.05$ for most coronal loops (Bray et al. 1991). Therefore, the critical plasma density around the looptop of typical coronal loops should be about $3.6\times10^{10}$ cm$^{-3}$ when the temperature is about $2\times10^{6}$ K and magnetic field strength is about 50 Gs. This value is consistent with the typical temperature in flaring loops.

Actually, it is difficult to give an exact description of the development of ballooning instability in a coronal plasma loop. Here, it is useful to estimate the characteristic time of the ballooning instability development. From the work of Shibasaki (2001), this characteristic time can be expressed,

\begin{equation}
t_{b}\approx\frac{2R}{C_{s}\sqrt{\beta_{c}}}\approx1.7\times10^{-2}\frac{R}{\sqrt{T\beta_{c}}}.
\end{equation}

Here, $C_{s}=\sqrt{\frac{\gamma p}{\rho}}=\sqrt{\frac{\gamma k_{B}T}{m_{i}}}\approx117.3\sqrt{T}$ is the sound speed. $\gamma$ is the ratio of specific heat capacity, usually $\gamma=\frac{5}{3}$ in the ideal gas. $m_{i}$ is the mass of ion. Considering a coronal loop with R = 25 Mm, the characteristic time of the ballooning instability development is about 17 minutes. Typically, $t_{b}\sim 100 - 1000$ s. This implies that the ballooning instability has a 10 - 20 minutes preflare developing process before the flaring loop eventually erupts.

The comparisons among the five timescales show the following relations:

\begin{equation}
t_{mc}\ll t_{c}\ll t_{f}<t_{b}\ll t_{d}.
\end{equation}

Equation (15) indicates: (1) MGP model is valid even in the photosphere, chromosphere as well as in corona, (2) the solar plasma loop have enough time to be heated by MGP process, and (3) the ballooning instability has a relatively short developing process before the loop eventually erupts.

\begin{figure}[ht]      
\begin{center}
   \includegraphics[width=12 cm]{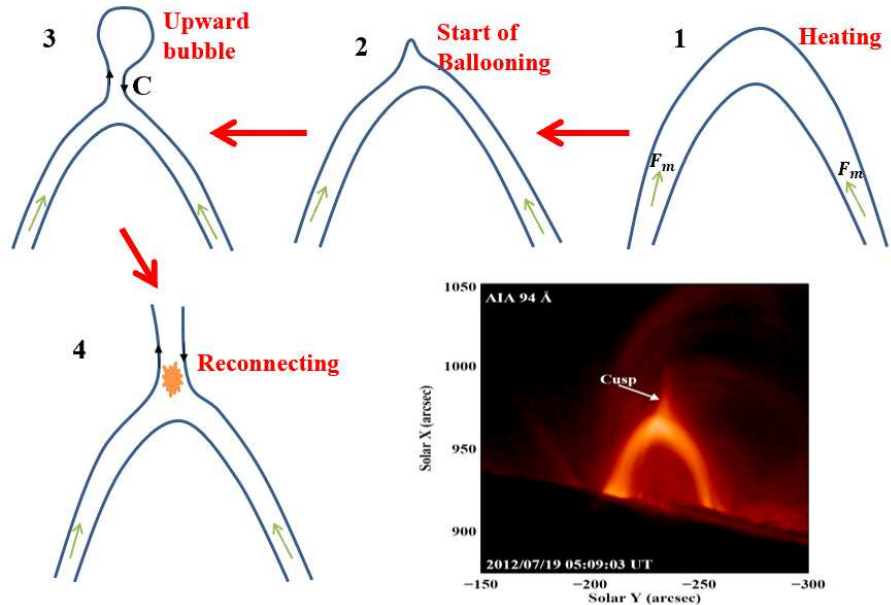}
\caption{The evolution of a solar plasma loop driving by MGP mechanism. 1. Heating of the plasma loop by MGP process. 2. Starting of ballooning instability and appearing of the finger structure. 3. Formation of the upward bubble and a X-point (C) of the magnetic field. 4. Magnetic reconnection and the formation of cusp-like configuration. The last panel is an example of bright hot cusp-configuration on the image of 94 \AA~ observed by AIA/SDO in the early phase of a M7.7 flare on 19-Jul-2012.}
\end{center}
\end{figure}

The whole process from the preflare heating of the plasma loop via ballooning instability around the looptop to erupting in cusp-like flare can be showed in Fig.4. Here, MGP process heats the looptop (marked 1), and ballooning instability will be triggered to start around the looptop when $\beta\geq\beta_{c}$, forms finger structures (marked 2). In a longitudinally homogeneous plasma loop or a long straight cylindrical plasma tube, the finger structures due to the ballooning instability may appear in the random direction around the loop. It is difficult to know where and which direction the finger structure will appear. However, as the solar plasma loops are not longitudinally homogeneous, the looptop is the weakest point of magnetic confinement for its weakest magnetic field strength and the highest plasma temperature, therefore, the finger structure due to the ballooning instability will happen first around the looptop and at the direction of upward. The threshold can be obtained from Equation (13). After the formation of finger structure, due to the continues injection of the pumping particles by the MGP process, the finger structure will expand and develop into an upward moving bubble and make the opposite magnetic field lines close to each other around C point beneath the bubble (marked 3). When the bubble rises to a certain height, a current sheet and an X-point will generate and trigger the magnetic reconnection above the looptop (marked 4). Finally the magnetic field lines will be broken, reconnect and release magnetic energy and energetic particles rapidly, and form cusp-like flares. The magnetic reconnection can also accelerate the charged particles and generate nonthermal particles violently. Therefore, the ballooning instability should be a result of MGP process, and it might be a precursor of the magnetic reconnection. The looptop becomes hot during the process of ballooning instability which may last for about 10-20 minutes in loops with radius of 25 Mm. Finally, a hot cusp-configuration can be observed.

The right bottom panel of Fig. 4 presents an observing EUV images of a solar flare the early phase, which is an example of hot cusp-like structure obtained at 94 \AA~ by AIA/SDO (Lemen et al. 2012) at 05:09:03 UT on 19-Jul-2012, just at the start of an M7.7 flare (Sun et al. 2014, Huang et al. 2016). Here, the looptop is much brighter than other parts. The breakup of the flaring loop primarily takes place around the looptop, and then it finally develops into cusp-like flare (Masuda et al. 1994, Masuda et al. 1995, Shibata et al. 1995, Karlicky et al. 2006).

In the previous literatures, Shibasaki (2001) and Hollweg (2006) also mentioned the role of magnetic-gradients in plasma loops and the possibly high-beta disruption triggered by the ballooning instability. They proposed that magnetic-gradient force would push the whole plasma as a fluid toward weak magnetic field region, and the magnetic field played as a converter of thermal random motion into coherent flow motion and instability. However, there are two distinct differences between our MGP model and the regime of Shibasaki (2001) and Hollweg (2006) (hereafter, simply say S-H regime): (1) S-H regime has not considered the dependence between magnetic-gradient force and the kinetic energy of particles, and therefore their plasma flow has no temperature change. Our MGP model emphasized that the magnetic-gradient force is proportional to the particles' kinetic energy. The higher the kinetic energy, the stronger the magnetic-gradient force acting on the charged particle, and therefore it will escape more easily from the lower atmosphere. (2) S-H regime does not include the solar gravitational force which is a key factor in MGP model. It was just because of the solar gravitational force to divide all particles into two groups: pumping particles and confined particles, they have different behaviors in the solar plasma loops.

\section{Application to the other solar phenomena}

The MGP model can be also applied to demonstrate other astrophysical processes, such as coronal heating (Tan 2014), the formation of the cold in sunspot near photosphere and the hot above it, the coronal plasma jets, and the fast solar wind above the coronal holes, etc. In this section, we try to apply the MGP model to provide a new explanation of some of the above phenomena.

\subsection{Sunspot}

It is well-known that sunspots are colder than the surrounding photosphere. The previous models explain this phenomenon as the strong magnetic fields of sunspots suppress the convective flows beneath the photosphere. However, a large number of observations indicate that the plasmas high up in the atmosphere above sunspots are always hotter than the surrounding chromosphere and corona at the same height, and this is the reason why the most solar flares always take place somewhere above sunspot active regions. Now that the strong magnetic suppression holds back the hot materials flowing into the region of sunspots from the solar interior, why the upper part is hot above the sunspot? The magnetic suppression hypothesis is hard to make a perfect demonstration on this phenomenon. Here, we try to apply MGP model to present a new explanation of the cold in sunspot and the hot above it.

As we know, the magnetic field is approximately vertical to the solar surface in and above the sunspots. Therefore, the Equation (5) should be changed into the following form,

\begin{equation}
F_{t}=-G_{B}\cdot\epsilon_{t}+mg(h).
\end{equation}

The starting energy should be a function of the height ($h$) above the solar surface,

\begin{equation}
\epsilon_{t0}(h)=mg(h)L_{B}(h).
\end{equation}

At each height, the charged particles with kinetic energy of $\epsilon_{t}>\epsilon_{t0}(h)$ will fly away and move upward, while the charged particles with kinetic energy of $\epsilon_{t}<\epsilon_{t0}(h)$ will stay beneath this height. Only the charged particles with energy around $\epsilon_{t0}(h)$ will stay around the height of $h$. Therefore, the starting energy ($\epsilon_{t0}$) should be an indicator of the temperature ($T$) at the height of $h$ above the sunspot.

In the region at the height of 0.02 - 10 $R_{s}$ above the sunspot, we may still adopt Equation (1) and (2) to express the distribution of magnetic field strength and its gradient approximately, then we may obtain a roughly estimation of the temperature distribution,

\begin{equation}
T(h)\approx2.2\times10^{-2}\frac{h}{(1+\frac{h}{R_{s}})^{2}}.
\end{equation}

However, in the region below the height of 0.02 $R_{s}$, Equations (1) and (2) are not suitable to describe the magnetic field and its gradient, we can not estimate the temperature distribution by using Equation (18). We have to adopt some modeling results of the magnetic field and its gradient to estimate the temperatures very close to the sunspots.

We assume that the solar photosphere has the averaged temperature ($T_{0}$) at about 5780 K and density of $10^{^{22}}m^{-3}$. When a sunspot appears, the magnetic field and its gradient will occur simultaneously. Under the joint-action of magnetic-gradient force and the solar gravitational force, the confined particles will stay near the sunspot for their low energy, while the energetic pumping particles will flow upward and carry a fraction of kinetic energy to the upper atmosphere. The temperature of the sunspot ($T_{s}$) will decrease for losing part of kinetic energy, which can be estimated by,

\begin{equation}
T_{s}=\frac{k_{B}T_{0}-E(\epsilon_{t}>\varepsilon_{t0})}{k_{B}N(\epsilon_{t}\leq\varepsilon_{t0})}\approx T_{0}-\frac{E(\epsilon_{t}>\varepsilon_{t0})}{k_{B}N(\epsilon_{t}\leq\varepsilon_{t0})}.
\end{equation}

Here, $f(\epsilon_{k})$ is supposed to be a Maxwellian distribution function which is dominated by temperature. $E(\epsilon_{t}>\epsilon_{t0})=\int_{\epsilon_{t0}}^{\infty}
f(\epsilon_{k})\epsilon_{k}d\epsilon_{k}$ is the kinetic energy carried by pumping particles. $N(\epsilon_{t}\leq\varepsilon_{t0})=\int_{0}^{\epsilon_{t0}}
f(\epsilon_{k})d\epsilon_{k}$ is the density of the confined particles in sunspot. When we suppose that $L_{B}\sim900$ km, then $T_{s}\approx4474$ K, 1306 K of temperature decrease from the initial state (Fig.5). This result is very close to the result of observations and simulations (Vernazza et al. 1981, etc.).

\begin{figure*}[ht] 
\begin{center}
   \includegraphics[width=10 cm]{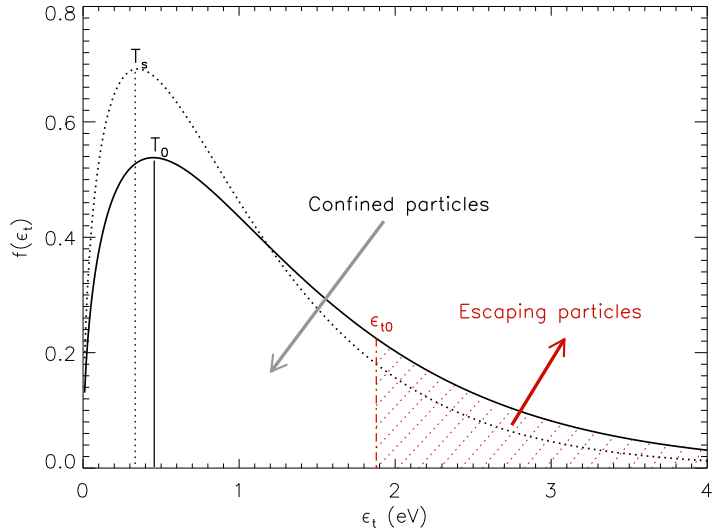}
\caption{The evolution of distribution functions of the plasmas around sunspot acting on MGP mechanism. $\epsilon_{t0}$ is the pumping starting energy. The black solid curve is the distribution function of the initial state of the photosphere ($T_{0}=5780$ K, $n_{n}=10^{22}m^{-3}$, $L_{B}\sim900$ km). It will develop into the dotted curve after losing the energetic escaping particles (the right red shadow region) and the temperature decreases to $T_{s}=4474$ K.}
\end{center}
\end{figure*}

\begin{figure*}[ht] 
\begin{center}
   \includegraphics[width=10 cm]{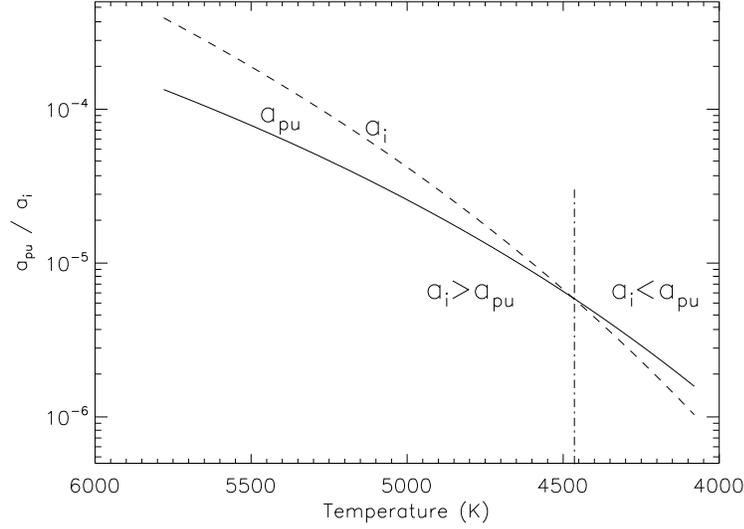}
\caption{The comparison between the pumping rate ($a_{up}$, solid line) and the degree of ionization ($a_{i}$, dashed line) at different temperature. The initial density is $n_{n}=10^{22}m^{-3}$.}
\end{center}
\end{figure*}

As we mentioned in Section 3.1, the magnetic field scale length should be about 2000 km around the sunspots, then Equation (18) leads to $T_{s}\approx5770$ K, only 10 degrees of temperature decrease from the initial state ($T_{0}=5780$ K, $n_{n}=10^{22}m^{-3}$). Here, we should realize that the MGP is a continuous process. The initial temperature of the photosphere is at 5780 K. It will decrease to 5770 K under the action of MGP process. Because the atmosphere still contains considerable charged energetic particles ($a_{i}=1.2\times10^{-4}$) which can be pumped to move upward by the magnetic-gradient force. Therefore, the temperature will be continuous to decrease. However, the degree of ionization will sharply decrease when the temperature decreases. Fig. 6 presents the comparison between the pumping rate and the degree of ionization at different temperature. It shows that the degree of ionization is larger than the pumping rate ($a_{i}>a_{up}$) when the temperature $T>4460$ K which implies that there are enough charged particles to be pumped by the MGP process. However, the degree of ionization is smaller than the pumping rate ($a_{i}<a_{up}$) when the temperature $T<4460$ K, which implies that there are no enough charged particles to be pumped to move upward in this temperature range. The minimum temperature of sunspot is about 4460 K.

\subsection{Solar plasma jets}

In solar atmosphere, plasma jets are ubiquitous in columnar collimated, beam-like eruptions that are magnetically rooted in the photosphere and shoot up along large-scale unipolar guide field reaching high into the corona. Solar plasma jets include spicules, H$\alpha$ surges, photospheric jets, chromospheric jets, coronal EUV and X-ray jets, and white-light polar jets (Moore et al. 2010). They represent important manifestations of ubiquitous solar transients especially onside coronal holes and their long periphery, which may be the source of mass and energy input to the solar upper atmosphere and the solar wind. The observed velocities of solar plasma jets range from several decades to more than 500 km s$^{-1}$ with height from a few thousand km up to several solar radii. The lifetimes of coronal EUV jets ranged from about 5 to 70 min. There are typically two models to explain the formation of solar plasma jets: the the magnetic reconnection model and the nonstandard blowout model. Despite the major advances made on both observations and theories of solar plasma jet, so far, many questions are still not completely understood, including its nature, their triggers, evolution, and contribution to the coronal heating and acceleration of solar wind (Raouafi et al. 2016). For example, the magnetic reconnection model can explain the formation of plasma jets related to solar eruptions in active regions, but it is difficult to explain why the velocity increases after the jet leave from its formation site, such as the type II spicules (De Pontieu et al. 2009, 2011, Samanta et al. 2019), the hot plasma ejections along the ultrafine magnetic channels from the solar surface upward to the corona (Ji et al. 2012), and polar jets, etc.

Here, we attempt to apply MGP mechanism to demonstrate the formation of the type II spicules, polar jets, and other solar plasma jets without relationships to solar activities. We assume that the escaping energetic particles in solar open magnetic configurations may form the upflow of plasma jets, the averaged velocity of the escaping particles can be an estimation of the velocity of upflow jet, which can be calculated,

\begin{equation}
v_{up}=\frac{\int_{\epsilon_{t0}}^{\infty}
f(\epsilon_{k})v_{\|}d\epsilon_{k}}{\int_{\epsilon_{t0}}^{\infty}
f(\epsilon_{k})d\epsilon_{k}}.
\end{equation}

Here, $v_{\|}\approx\sqrt{\frac{2\epsilon_{k}}{m}}$ is the vertical velocity component of the escaping particles. Because the starting energy $\epsilon_{t0}$ of the pumping particles is a function of the height above the solar surface (Equation 6), the velocity $v_{up}$ of upflow jet is also a function of the height.

\begin{figure*}[ht] 
\begin{center}
   \includegraphics[width=10 cm]{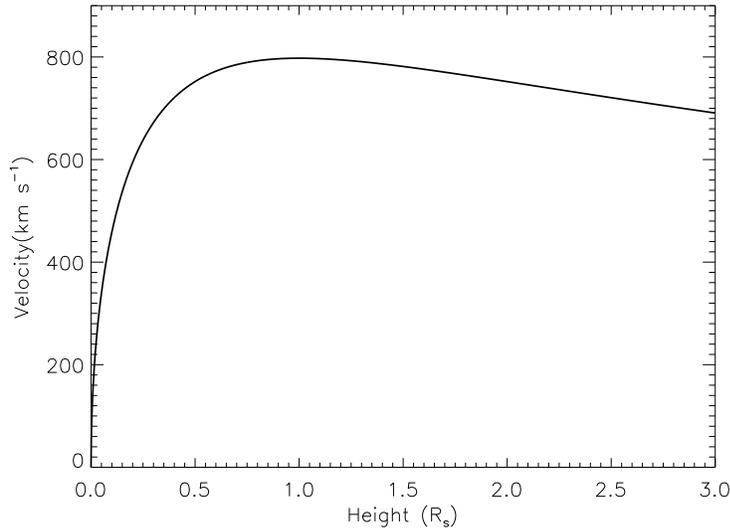}
\caption{The velocity of upflow at different height above the solar surface driving by MGP mechanism in an open magnetic configuration. Here, the unit of the height is solar radius $R_{s}$.}
\end{center}
\end{figure*}

If we assume the expressions of magnetic field and gradient are still valid in the form of Equation (1) and (2). Then we may obtain an approximated profile of the velocity above solar surface. Fig. 7 presents the velocity of upflows at different height above the solar surface driving by MGP mechanism in an open magnetic configuration: $v_{up}\sim$ 20 - 30 km s$^{-1}$ near the photosphere, $v_{up}\sim$ 40 - 60 km s$^{-1}$ in the chromosphere, $v_{up}\sim$ 150 - 200 km s$^{-1}$ at the bottom of corona, and $v_{up}\sim$ 800 km s$^{-1}$ in the corona at height of 1.0 $R_{s}$. The velocity is slightly decreasing beyond the height of 1.0 $R_{s}$. These results are nearly in line with the observations (Savcheva et al. 2007).

It is possible that Equation (1) and (2) are not valid exactly for describing the magnetic field and gradient beyond active regions. However, the above estimation still implies that MGP mechanism may provide a reasonable explanation for the formation of solar plasma jets. One of the advantages is that the MGP mechanism can explain the velocity increasing of the solar plasma jets from the solar photosphere to high corona after they leave from their source region.

Similarly, because the compact bodies, such as the white dwarfs, neutron stars, and black holes also have strong magnetic field and the related magnetic-gradient in their surrounding atmosphere, the magnetic-gradient force can also drive and form fast plasma jets. Equation (17) indicates that the starting energy is proportional to the gravitational force: $\epsilon_{t0}\propto g(h)$. Because the compact bodies have much more strong gravitation force, and therefore the starting energy is much higher than that in solar conditions. Additionally, the atmosphere around the compact bodies is much hotter than the solar atmosphere. All these facts imply that the plasma jets may much more high speeds.

\subsection{Fast solar wind above coronal holes}

The solar wind is a stream of charged particles (including electrons, protons, and $\alpha$ particles, etc.) released from the solar upper corona. Among them, the fast solar wind has a flow speed exceeding 200 - 300 km s$^{-1}$ at 2 - 3 $R_{s}$, near 700 - 800 km s$^{-1}$ well below 10 $R_{s}$. The fast solar winds are believed to originate from the coronal holes, which are funnel-like regions of open magnetic fields (Bravo \& Stewart 1997, Wilhelm et al. 2000).

Obviously, there exists magnetic-gradient in the coronal hole and the funnel-like regions of open magnetic field. Reasonably, we may apply the energetic particle flows driven by magnetic-gradient force to explain the formation of fast solar wind similar to the solar plasma jets. It was just the escaping energetic particles pumped by the magnetic gradient force from coronal hole and the funnel-like regions of open magnetic field to form the fast solar wind. We may approximately adopt Equation (20) to estimate the flow speed of fast solar wind at different heights. Here, the magnetic field and its gradient are unknown. We may try to assume its magnetic field scale height ($L_{B}$) reasonably. For example, at height of 2 - 3 $R_{s}$ above solar surface, $T_{0}\sim10^{6}$ K, $L_{B}\sim 10^{6}$ km, then $v_{up}\sim$ 350 km s$^{-1}$. At height of 10 $R_{s}$ above the photosphere, $T_{0}\sim10^{6}$ K, $L_{B}\sim10^{7}$ km, then $v_{up}\sim$ 730 km s$^{-1}$. These values are well in accordance with observations (Feldman \& Landi 2005). This estimation also indicates that the fast solar winds are possibly the energetic particle flow driven by magnetic-gradient force above coronal holes.

Obviously, here we need a more exact estimation of the magnetic fields in the high corona from 2-3 $R_{s}$ to beyond 10 $R_{s}$. And this requires multiple diagnostic tools of coronal magnetic fields.

\section{Conclusions}

In summary, we obtain the following conclusions from this work:

(1) The calculations and comparisons between the collision timescales ($t_{c}(ia)$ and $t_{c}(ii)$), the magnetic cyclotron period ($t_{mc}$), and the lifetime of solar plasma loops indicate that the MGP model is valid even in the solar photosphere, chromosphere, as well as in the corona.

(2) The MGP process can heat the top region of solar plasma loops up to several million Kelvin, make the looptop exceeds the critical $\beta_{c}$, trigger the ballooning instability to produce finger structures, result in the looptop expanding, plumping, out of shape, and produce an upward bubble and reversed magnetic field, and finally trigger the eruption around a cusp-like structure. Therefore, the MGP process can provide a natural driver of the solar eruptions.

(3) The MGP model can be applied to demonstrate the low temperature of sunspot, the formation of solar plasma jets, type-II spicule, and fast solar wind above the coronal holes.

The magnetic-gradient force drives energetic particle upflows in solar plasma loops, extract the kinetic energy from the underlying plasma, convey and transport them to the upper corona. This mechanism provides an natural approach to explain the processes occurring in the early phase of solar eruptions, including transporting the energetic particles and kinetic energy to the top region of plasma loops, increasing the temperature, pressure and plasma beta, stimulating and triggering the ballooning instability, and finally driving eruptions. Here, the magnetic-gradient plays a key role in the converting of not only the energetic particles but also the kinetic energy for solar eruptions. The pumping particle flows play a crucial role of energy storage in corona by carrying the energetic particles to pile up around the looptops. When the looptop becomes overpressure, it departs from the equilibrium and trigger the ballooning instability near the looptop and generate the cusp-like flare eruptions. This is a natural triggering mechanism of solar eruptions, which does build a direct connections among the erupting energy, the underlying plasma's motions, and the magnetic configurations. This mechanism implies that the released energy during the eruption primarily comes from the solar interior, and their transporting channels are magnetized plasma loops.

The above deductions suggest that the magnetic-gradient possibly dominate the occurrence of solar eruptions. It is very meaningful for solar activity prediction to diagnose the magnetic field and its structures in solar atmosphere. In open magnetic flux loop, the energetic particle flows driven by magnetic-gradient force may provide a possible demonstration of the the plasma jets and fast solar wind. Furthermore, it will also help us the understand the fast jets related to black holes or other compact celestial bodies, such as neutron stars and black holes.

The energetic particle flows driven by magnetic-gradient force are fundamental and ubiquitous phenomena in the inhomogeneous magnetized plasmas. The energetic particles will be driven to move to the weak magnetic field region and make the plasma temperature to become more and more uneven. Consequently, this will excite plasma instabilities, such as the ballooning instability. This furthermore results in the release of energetic particles and energy. It is possibly that it was just the magnetic-gradient force trigger and drive the generation of the major disruption in Tokamak plasmas. The MGP mechanism might give us a bit of enlightenments for controlling the nuclear fusion plasmas and help us to understand the formation of various astrophysical plasma jets.

\begin{acknowledgements}
The authors would thank the referee for his helpful and valuable comments to improve the manuscript of this paper. This work adopted EUV observation obtained by AIA/SDO, soft X-ray observation by GOES satellite. It is supported by NSFC Grants 11433006, 11573039, 11661161015, 11790301 and 11973057.
\end{acknowledgements}

\label{lastpage}

\end{document}